\documentclass[aps,prl,showpacs,twocolumn,floats]{revtex4}

\usepackage{graphicx}
\usepackage{dcolumn}
\usepackage{amsmath}
\usepackage{bm}

\begin{document}

\newcommand{\ben}{\begin{equation}}
\newcommand{\een}{\end{equation}}
\newcommand{\bena}{\begin{eqnarray}}
\newcommand{\eena}{\end{eqnarray}}

\title{Macroscopic quantum effects generated by the acoustic wave in a molecular magnet}

\author{Gwang-Hee Kim$^1$ and  E. M. Chudnovsky$^2$}
\affiliation{\it{${}^{1}$Department of Physics, Sejong University,
Seoul 143-747, Republic of Korea \\
${}^{2}$Department of Physics and Astronomy, Lehman College, City
University of New York, 250 Bedford Park Boulevard West, Bronx, New
York 10468-1589, USA}}

\begin{abstract}
We have shown that the size of the magnetization step due to
resonant spin tunneling in a molecular magnet can be strongly
affected by sound. The transverse acoustic wave can also generate
macroscopic quantum beats of the magnetization during the field
sweep.

\end{abstract}


\maketitle


Single-molecule magnets(SMMs) have attracted much interest because
they provide possibility to observe quantum effects at the
macroscopic scale. Among these effects are step-wise magnetization
curve caused by resonant spin tunneling
\cite{Friedman,EC-JT-book}, topological interference of tunneling
trajectories \cite{Wern-Sessoli}, and crossover between classical
and quantum superparamagnetism \cite{Theory,Exp}. Landau-Zener
theory has been used to describe spin transitions that occur
during the field sweep \cite{LZ-book}. It has been recognized that
spin-phonon interactions play an important role in the dynamics of
spins in molecular magnets \cite{Gar-Chu-97,UnivDec}. Possibility
of Rabi oscillations of spins caused by the acoustic wave has been
studied \cite{Cal-Chu}. In recent years the effect of sound on
molecular magnets has been explored in experiment \cite{Tejada}.
In this Letter we show that sound can significantly affect the
size of the magnetization step due to resonant spin tunneling. In
the presence of the field sweep an acoustic wave can also generate
quantum beats of the magnetization of a macroscopic sample. We
compute the parameters of the sound that are necessary to observe
these effects.

\begin{figure}
\vskip -0.5cm
\includegraphics[width=9.0cm]{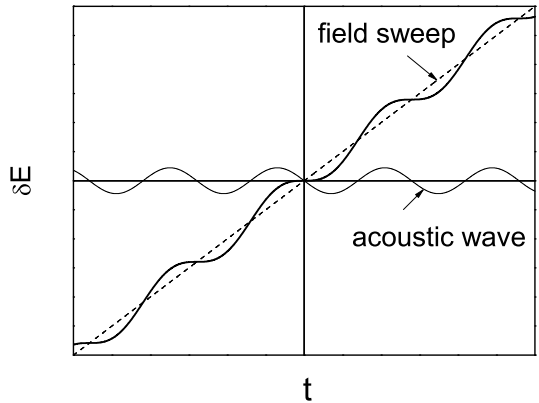}
\vskip -0.5cm \caption[0]{Schematic diagram of the time dependence
of the distance between spin energy levels. Thin solid line: The
effect of the acoustic wave without the field sweep. Thick solid
line: Field sweep is modulated by the acoustic
wave.}\label{fig:ene-r0.02}
\end{figure}
The effect we are after is illustrated in Fig. 1. The lower curve
shows how the sound modulates the distance between spin energy
levels in the absence of the field sweep. In this case the phase
of Rabi oscillations caused by a propagating sound wave depends on
coordinates such that the oscillations average out over the volume
of the sample if the latter is large compared to the wavelength of
sound. In the presence of the field sweep (the upper curve) the
phase of the Rabi oscillations is still a function of coordinates.
However, the Landau-Zener probability of spin transitions that
contribute to the oscillations depends on the rate of the field
sweep. That rate becomes modulated by the sound. Consequently, the
regions of the sample that contribute most to the dynamics of the
magnetization add their contributions constructively. The
resulting oscillations of the magnetic moment of the sample can be
observed in a macroscopic experiment.

We consider a crystal of single-molecule magnets with the
Hamiltonian,
\begin{eqnarray}\label{H}
{\cal{H}_{\rm SMM}}=-D S^{2}_{z}-g \mu_B H_z S_z +{\cal{H}_{\rm
trans}} \,,
\end{eqnarray}
where $S_{i}$ are Cartesian components of the spin operator and
$D$ is the second-order anisotropy constant. The second term is
the Zeeman energy due to the longitudinal field $H_z$, with $g$
being the gyromagnetic factor and $\mu_B$ being the Bohr magneton.
The last term includes the transverse magnetic field and the
transverse anisotropy, which produce level splitting. Local
rotation produced by a transverse acoustic wave of frequency
$\omega=c_tk$, wave vector $k$, and amplitude $u_0$, polarized
along the $y$ axis and running along the $x$ axis, is given by
\cite{LL}
\begin{eqnarray}
{\delta {\bm \phi}({r})}={1 \over 2} ku_0 \cos(kx-\omega t)
\hat{z} \,. \label{angle}
\end{eqnarray}
Due to the rotation of the local anisotropy axis by sound, the
spin Hamiltonian becomes \cite{UnivDec}
\begin{eqnarray}
{\cal{H}}=e^{-i{\delta {\bm \phi}}\cdot {\hat {\bf S}}}
{\cal{H}_{\rm SMM}} e^{i{\delta {\bm \phi}}\cdot {\hat {\bf S}}}
\,.
\end{eqnarray}
The simplest solution of the problem for an individual spin can be
obtained in the coordinate frame that is rigidly coupled to the
local crystallographic axes. The wave functions in the laboratory
and lattice frames, $|\Psi \rangle$ and $|\Psi^{(\rm lat)}
\rangle$, are related through
\begin{equation}
|\Psi^{(\rm lat)} \rangle= e^{i{\delta {\bm \phi}}\cdot {\hat {\bf
S}}} |\Psi \rangle\,,
\end{equation}
while the spin Hamiltonian in the lattice-frame is given by
\cite{EC-94,EC-Martinez,UnivDec}
\begin{eqnarray}
{\cal {H}}^{(\rm lat)}={\cal {H}}_{\rm SMM}- \hbar \hat{\bf S}
\cdot {\bf \Omega} \,, \label{hamil-lat}
\end{eqnarray}
with
\begin{eqnarray}
{\bf \Omega} \equiv \delta \dot{\bm \phi} ={\omega^2 \over 2 c_t}
u_0 \sin(kx-\omega t) \hat{z}\,.
\end{eqnarray}
We are going to solve the problem locally for each spin in the
lattice frame and then use the above formulas to obtain the
solution for the entire crystal in the laboratory frame.

\begin{figure}
\vskip -0.5cm
\includegraphics[width=9.0cm]{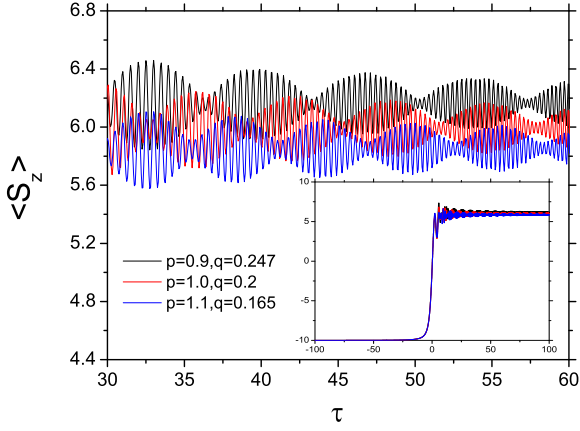}
\vskip -0.5cm \caption[0]{Color online: Time dependence of the
average spin of the sample for given values of $p$ and $q$ at
$S=10$, $M=0$, and $\gamma=0.02$. Inset: Magnetization step and
oscillations in the wider range of
$\tau$.}\label{fig:beat-m0-r0.02}
\end{figure}
In the absence of transverse terms the energy levels of the
Hamiltonian (\ref{H}) are
\begin{equation}
E_m = -D m^2 - g \mu_B H_z m\,,
\end{equation}
where $\hat{S}_z |m \rangle=m |m \rangle$. Close to the resonance
between $|-S \rangle$ and $|S-M \rangle$ the Hamiltonian
(\ref{hamil-lat}) can be projected onto these states, resulting
effectively in a two-level model:
 \bena  {\cal{H}}^{(\rm lat)}_{\rm eff}= -{1 \over 2} \Delta \hat{\sigma}_x
 - \delta E(\hat{\sigma}_z + \hat{I} ) , \label{eff-hamil}
 \eena
where
\begin{eqnarray}
& & \delta E= \left(S-{M \over 2} \right) \left[g \mu_B ct+{\hbar
\omega_R \over S} \sin(kx-wt)\right]
\nonumber \\
& & \hat{\sigma}_z=|S-M \rangle \langle S-M|-|-S \rangle \langle
-S| \nonumber \\
& &  \hat{\sigma}_x =|S-M \rangle \langle-S|+|-S \rangle \langle
S-M| \nonumber \\
& & \hat{I}=|S-M \rangle \langle S-M|+|-S \rangle \langle -S|\,,
\end{eqnarray}
$\Delta$ is the splitting of the resonant levels, $c=dH_z/dt$ is
the field sweep rate, and
 \bena
\omega_R={\omega^2 \over 2 c_t} u_0 S   \eena
is the Rabi frequency. Treating $x$ as a parameter, we express the
corresponding wave function as
\bena |\Psi^{\rm (lat)}_{\rm eff}(t)\rangle =b_{S-M}(t)|S-M
\rangle+b_{-S}(t)|-S \rangle. \eena and solve the time-dependent
Schr\"odinger equation,
\bena i \hbar {\partial |\Psi^{\rm (lat)}_{\rm eff}(t)\rangle \over
\partial t} ={\cal{H}}^{(\rm lat)}_{\rm eff} |\Psi^{\rm (lat)}_{\rm eff}(t)\rangle
\label{schro}\,, \eena
that at $M = 0$ becomes equivalent to the following two coupled
differential equations:
\bena {d b_{ S} \over d \tau}&=&2iS \left[\gamma \tau -{q p \over
S} \sin ( p  \tau -{k x}) \right] b_{S}+{i \over 2} b_{-S}
\nonumber
\\
{d b_{-S} \over d \tau}&=& {i \over 2} b_{S}, \label{diff-both}
\eena
where we introduced dimensionless
\begin{equation}
\tau=t\left(\frac{\Delta}{\hbar}\right), \quad \gamma =
\frac{\hbar g \mu_B c}{ \Delta^{2}}, \quad
p=\frac{\hbar\omega}{\Delta}, \quad q=\frac{\omega_R}{\omega}\,.
\end{equation}
We consider samples of length that is large compared to the
wavelength of the sound. The expectation value of the
$z$-projection of the spin at $M=0$ is given by
\bena \langle \Psi(t) | \hat{S}_z | \Psi (t) \rangle = S \left|
b_{S}(\tau)\right|^2+(-S)\left| b_{-S}(\tau)\right|^2\,. \eena

\begin{figure}
\vskip -0.5cm
\includegraphics[width=9.0cm]{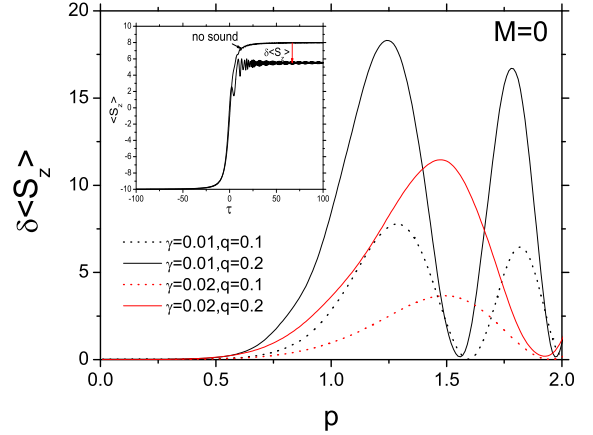}
\vskip -0.5cm \caption[0]{Color online: Final magnetization vs $p$
for the $M=0$ step with and without acoustic wave at different
$\gamma$ and $q$. Inset: $\langle S_z \rangle$ vs $\tau$  at
$\gamma=0.01$, $q=0.2$ and $p=0.9$.}\label{fig:diff}
\end{figure}

Eqs. (\ref{diff-both}) have been solved numerically.  Fig.
\ref{fig:beat-m0-r0.02} illustrates situation when the field was
changing at a constant rate $\gamma$ and a pulse of sound was
introduced shortly before reaching the resonance between the
$|-S\rangle$ and $|S\rangle$ states. The tunnel splitting is
assumed to be sufficient to produce transitions between these two
states. The most striking feature of the magnetization dynamics
observed in simulations are the beats which are in line with the
idea outlined in the introduction. Fig. \ref{fig:diff} shows the
$p$-dependence of the final magnetization on crossing the step for
various values of $\gamma$ and $q$. This strong  dependence of the
magnetization step on frequency and amplitude of the acoustic
wave, as well as on the sweep rate, is one of our main results. We
believe that it should not be difficult to observe this effect in
experiment.

\begin{figure}
\vskip -0.5cm
\includegraphics[width=9.0cm]{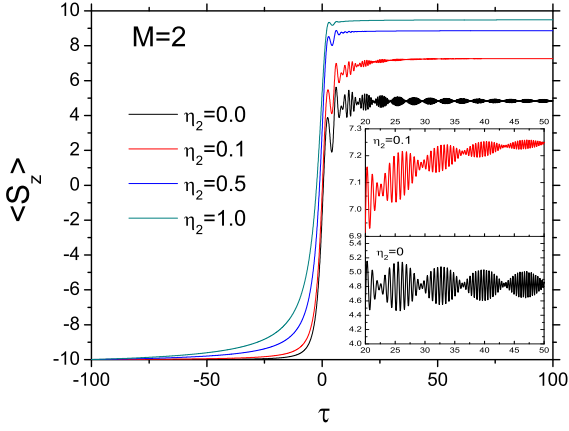}
\vskip -0.5cm \caption[0]{Color online: $\langle S_z \rangle$ vs
$\tau$ for $\eta_2 =$ 0, 0.1, 0.5, and 1.0 at $S=10$, $p=0.9$,
$q=0.247$, $\gamma=0.02$, and $M=2$. Inset: Oscillations between
$\tau=20$ and $\tau=50$ at $\eta_2=$0 and 0.1. Note that the beats
and the oscillatory behavior disappear as $\eta_2$
increases.}\label{fig:sz-m2-relax}
\end{figure}
\begin{figure}
\vskip -0.5cm
\includegraphics[width=9.0cm]{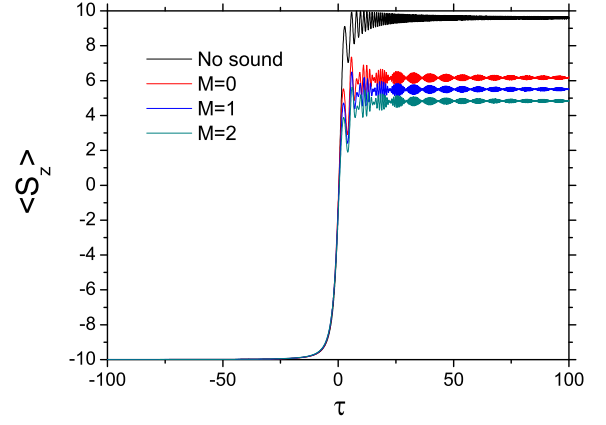}
\vskip -0.5cm \caption[0]{Color online:$\langle S_z \rangle$ vs
$\tau$ for $M=$ 0, 1, and 2, at $\eta_M=0$, $S=10$, $p=0.9$,
$q=0.247$, and $\gamma=0.02$. }\label{fig:sz-m=var}
\end{figure}

Another possible experimental situation corresponds to the sample
initially saturated in the $|-S\rangle$ state, after which the
acoustic power of frequency $\omega \approx \Delta/\hbar$ is
applied to the crystal and maintained during the sweep. Here
$|S-M\rangle$ is the level that at a given sweep rate provides
significant probability of the transition when it is crossed by
the $|-S\rangle$ level. In order to study such a problem, we need
to know the rate of relaxation of the $|S-M \rangle$ state to the
lower energy states. Defining $\Gamma_{S-M+1,S-M}$  as the rate of
the $|S-M \rangle \rightarrow |S-M+1 \rangle$ transition and
introducing $\eta_M=\hbar \Gamma_{S-M+1,S-M}/\Delta$, we obtain
two coupled differential equations:
\bena {d b_{ S-M} \over d \tau}&=& i z b_{S-M}+{i \over 2} b_{-S},
\nonumber
\\
{d b_{-S} \over d \tau}&=& {i \over 2} b_{S-M},
\label{diff-both-M} \eena
where
\begin{equation}
z = (2S-M) \left[\gamma \tau -(q p/S) \sin \left( p \tau -kx
\right) \right]+i\eta_M /2\,.
\end{equation}
As the lifetimes of the excited states with $|S-M+1
\rangle$,...,$|S-1 \rangle$ are shorter than the lifetime of $|S-M
\rangle$, their contributions to the above equations can be
neglected. Then
\bena \langle S_z \rangle=-2S |b_{-S}(\tau)|^2-M |b_{S-M}(\tau)|^2
+S \label{sz-M-red} \eena
We solve Eqs.(\ref{diff-both-M}) numerically for selected values of
$\gamma$ and $\eta_M$. In the overdamped case, $\Gamma_{S-M+1,S-M}
\gg \Delta$, we find no Rabi oscillations. For the underdamped case,
$\Gamma_{S-M+1,S-M} \ll \Delta$, the numerical solution is
illustrated in Fig. \ref{fig:sz-m2-relax}. As the damping increases,
the magnetization jump becomes more pronounced but the oscillatory
dynamics disappears. The comparative behavior of different
resonances is shown in Fig. \ref{fig:sz-m=var}.

Let us now study the optimal conditions for the observation of the
macroscopic acoustic Rabi effect studied above. Defining
$\omega_q(\tau,x) \equiv S [\gamma \tau -(q p/S) \sin(p\tau- 2 \pi
x/\lambda)]$ where $\lambda = 2\pi/k$ is the wavelength of the
sound,  the coupled differential equations (\ref{diff-both}) can
be written as
\bena \ddot{b}_S - 2 i \omega_q \dot{b}_S- \left(2 i
\dot{\omega}_q-{1 \over 4} \right) b_S=0\,, \eena
where $\dot{b}_{S}=d b_S/d \tau$, $\ddot{b}_{S}=d^{2} b_S/d
\tau^{2}$ and so on. Introducing $b_{S}=d_{S}\exp[i v(\tau)]$ and
selecting $\dot{v}=\omega_q$, we get
\bena \ddot{d}_S + \left(-i\dot{\omega}_q +\omega^{2}_{q}+{1 \over
4}\right) d_S=0\,, \eena
which describes damped oscillations.  At $q\neq 0$ we have, e.g.,
for $x=0$ and $x=\lambda/2$
\begin{eqnarray}
\omega_q (\tau,0)=S\left[ \gamma \tau - {q p \over S} \sin(p\tau)
\right],
\\
\omega_q(\tau, \lambda/2)=S\left[ \gamma \tau + {q p \over S}
\sin(p\tau) \right],
\end{eqnarray}
respectively.
This implies that each frequency generates slow and fast oscillatory
regions due to the sinusoidal function, and they show different
damped oscillatory structures in a given range of $\tau$. In other
words if $\omega_q (\tau,0)$ is larger than $\omega_0\equiv S \gamma
\tau$ in some range of $\tau$, $\omega_q (\tau,\lambda/2)$ is
smaller than $\omega_0 $, and vice versa (see Fig.
\ref{fig:ene-r0.02}). At $\tau \sim 0$ the frequencies are
approximately given by $\omega_q (\tau,0)\simeq S( \gamma  - q
p^2/S) \tau$ and $\omega_q(\tau, \lambda/2)\simeq S( \gamma + q
p^2/S) \tau$
%
%
Introducing $\tan\theta_1=\dot{\omega}_q (\tau,0)$,
$\tan\theta_2=\dot{\omega}_q (\tau,\lambda/2)$, and $w=qp^2 /S$, we
get
\begin{eqnarray}
|\tan(\theta_1-\theta_2)|=\left|{2 w \over
1+\gamma^{2}-w^2}\right|,
\end{eqnarray}
which increases monotonically in the range of $0<w<1+\gamma^{2}$.
Under this condition, let us first consider two limiting cases: $w
\gg \gamma$ and $w \ll \gamma$.  In the first case
$(\theta_1-\theta_2)$ increases with $w$, and thereby $|\omega_q
(\tau,0)-\omega_q(\tau, \lambda/2)|$ also increases, which is a less
favorable situation for the beats. In the second case we have
$\theta_1 \simeq \theta_2$, which results in $\omega_q(\tau,0)
\simeq \omega_q(\tau, \lambda/2)\simeq \omega_0$. This also is not a
favorable situation for the beats because it does not generate slow
and fast oscillatory regions for $x=0$ and $x=\lambda/2$, as
discussed previously. The optimal condition for pronounced beats is
then
\bena \gamma \simeq {q p^2 \over S}\,.  \label{optimal-beat} \eena

\begin{figure}
\vskip -0.5cm
\includegraphics[width=9.0cm]{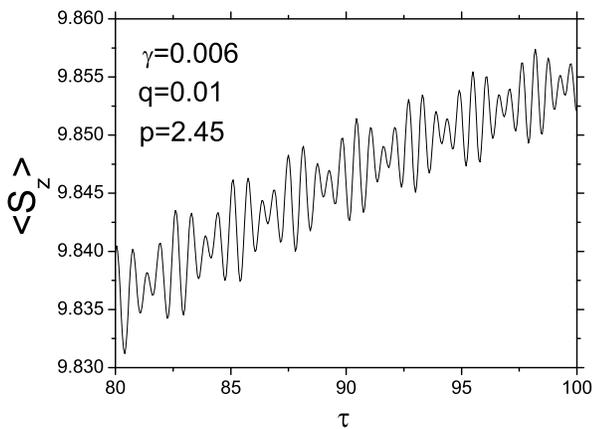}
\vskip -0.5cm \caption[0]{Time dependence of the magnetization of
the sample for $M=0$ at $S=10$, $\gamma=0.006$, $q=0.01$, and
$p=2.45$. }\label{fig:sz-m=0-exp}
\end{figure}

We shall now discuss  what the above condition means for
experiment. It is easy to see that Eq.\ (\ref{optimal-beat}) is
equivalent to
\begin{equation}
\frac{u_0}{\lambda} = \frac{q}{\pi S}\,.
\end{equation}
The validity of the continuous elastic theory that we employed
requires $u_0 \ll \lambda$, that is, one needs to satisfy the
condition $q < 1$. This is not sufficient, though. Since
experiments on molecular magnets require temperature in the kelvin
range or lower, one should also be concerned with the power of the
sound. It should be sufficiently low to avoid the unwanted heating
of the sample. The power per cross-sectional area of the sample is
given by $P/A = \frac{1}{2}\rho u_0^2\omega^2c_t$. For, e.g.,  the
parameters of Fe-8 molecular magnet we find that the optimal
conditions of the experiment require sound of frequency $f =
0.5$MHz -$ 1$MHz and power in the range
$100$W/cm$^{2}$-$200$W/cm$^{2}$ introduced into the sample
simultaneously with the field sweep of $1$kG/s. Time dependence of
the magnetization under these conditions is shown in Fig.
\ref{fig:sz-m=0-exp}.

In conclusion, we have demonstrated that the size of the
magnetization step due to resonant spin tunneling in molecular
magnets can be strongly affected by sound. The acoustic wave can
also generate macroscopic quantum beats of the magnetization
during a field sweep. The required frequency (MHz) and power
(0.1kW/cm$^2$) of the sound, and the required sweep rate (1kG/s)
are within experimental reach.

The authors are grateful to D. A. Garanin for useful discussions.
The work of GHK has been supported by the Grant No.
R01-2005-000-10303-0 from Basic Research Program of the Korea
Science and Engineering Foundation. The work of EMC has been
supported by the NSF Grant No. DMR-0703639.

\end{document}